\documentclass[letterpaper, 10 pt, conference]{ieeeconf}  

\IEEEoverridecommandlockouts                              

\usepackage{amsmath,amsthm,amssymb}
\usepackage{graphicx,fixltx2e}
\usepackage{hyperref}
\usepackage{xy}
\usepackage[]{algorithm2e}
\usepackage{bbm}
\usepackage{todonotes}
\usepackage{float}
\usepackage{cite}
\usepackage[utf8]{inputenc}
\usepackage[T1]{fontenc}

\newcommand{\N}{\mathbb{N}}

\newcommand{\R}{\mathbb{R}}

\newcommand{\mc}[1]{\mathcal{#1}}

\makeatletter
\newcommand{\pushright}[1]{\ifmeasuring@#1\else\omit\hfill$\displaystyle#1$\fi\ignorespaces}
\newcommand{\pushleft}[1]{\ifmeasuring@#1\else\omit$\displaystyle#1$\hfill\fi\ignorespaces}
\makeatother

\newenvironment{problem}[2][Problem]{\begin{trivlist}
\item[\hskip \labelsep {\bfseries #1}\hskip \labelsep {\bfseries #2.}]}{\end{trivlist}}

\newenvironment{proposition}[2][Proposition]{\begin{trivlist}
\item[\hskip \labelsep {\bfseries #1}\hskip \labelsep {\bfseries #2.}]}{\end{trivlist}}

\allowdisplaybreaks

\title{\LARGE \bf
Controller Synthesis for Discrete-time Hybrid Polynomial Systems via Occupation Measures
}

\author{Weiqiao Han and Russ Tedrake
\thanks{Computer Science and Artificial Intelligence Laboratory, Massachusetts Institute of Technology, 77 Massachusetts Avenue, Cambridge, MA 02139, USA. {\tt\small weiqiaoh,russt@mit.edu}}
\thanks{This work was supported by Air Force/Lincoln Laboratory Award No. 7000374874, and Department of the Navy, Office of Naval Research, Award No. N00014-18-1-2210. Any opinions, findings, and conclusions or recommendations expressed in this material are those of the authors and do not necessarily reflect the views of the Office of Naval Research.}%
}

\begin{document}

\newcounter{eqn}

\maketitle
\thispagestyle{empty}
\pagestyle{empty}

\begin{abstract}
We consider the feedback design for stabilizing a rigid body system by making and breaking multiple contacts with the environment without prespecifying the timing or the number of occurrence of the contacts.
We model such a system as a discrete-time hybrid polynomial system, where the state-input space is partitioned into several polytopic regions with each region associated with a different polynomial dynamics equation.
Based on the notion of occupation measures, we present a novel controller synthesis approach that solves finite-dimensional semidefinite programs as approximations to an infinite-dimensional linear program to stabilize the system.
The optimization formulation is simple and convex, and for any fixed degree of approximations the computational complexity is polynomial in the state and control input dimensions.
We illustrate our approach on some robotics examples.

\end{abstract}

\section{Introduction}
In robot locomotion and manipulation, it is common that a legged robot balances itself or a robotic hand manipulates an object by making and breaking multiple contacts with the environment.
However, there are no simple rules to stabilize such a system.
Local stabilization methods, such as linear-quadratic regulator (LQR), are unable to reason about the change of system dynamics.
In this paper, we model the system as a hybrid system, or more specifically a piecewise polynomial system, i.e., a system whose state-input space is partitioned into several polytopic regions, with each region associated with a different polynomial dynamics equation.
Such a hybrid system modeling approach is a natural fit for many problems in robotics.
For example, a system with contacts is hybrid, with each hybrid mode associated with a particular contact mode.

Controller synthesis for hybrid systems is a long-standing challenging problem.
The earliest work \cite{witsenhausen1966class} dates back to 1960s, where the hybrid model and the optimization formulation for controlling the system were introduced.
More recently, Branicky et al.  considered several algorithms for optimal control of hybrid systems \cite{branicky1995algorithms, branicky1995studies} and derived necessary conditions for the existence of optimal control law \cite{branicky1998unified}. 
Approaches based on dynamic programming \cite{branicky1998unified, hedlund2002convex, lygeros2008hybrid} and the maximum principle \cite{sussmann2000set, piccoli1999necessary, shaikh2002trajectory, shaikh2007hybrid} were proposed by several authors to solve the hybrid optimal control problem.
(More reviews on hybrid systems can be found in \cite{antsaklis2003hybrid, goebel2009hybrid, zhu2015optimal, lunze2009handbook, tabuada2009verification}.)
Approaches based on dynamic programming design control policies for the whole discretized state space and the computational complexity grows exponentially with respect to the dimension of the state.
Approaches based on the maximum principle are trajectory optimization methods, which design a trajectory for a specific initial state, and they usually assume mode switch sequences are fixed, while we are more interested in methods that automatically find mode switch sequences.
Recent works based on the hybrid minimum principle and gradients \cite{passenberg2010minimum,passenberg2010algorithm} do drop the assumption of fixed mode switch sequences.
Posa et al. \cite{posa2014direct} avoided the mode switch problem by formulating the linear complementarity problem for physical contacts.
However, we are interested in designing control policies that cover most of the state space rather than designing a trajectory for a fixed initial state.
Sampling based approaches \cite{branicky2003rrts,branicky2006sampling,sadraddini2018sampling} achieve a compromise between covering the state space with control policies and designing a single trajectory for an initial state, and are promising for high-dimensional systems.

\begin{figure}[t]
  \begin{minipage}{0.49\linewidth}
  \centering 
  \includegraphics[width=0.5\linewidth]{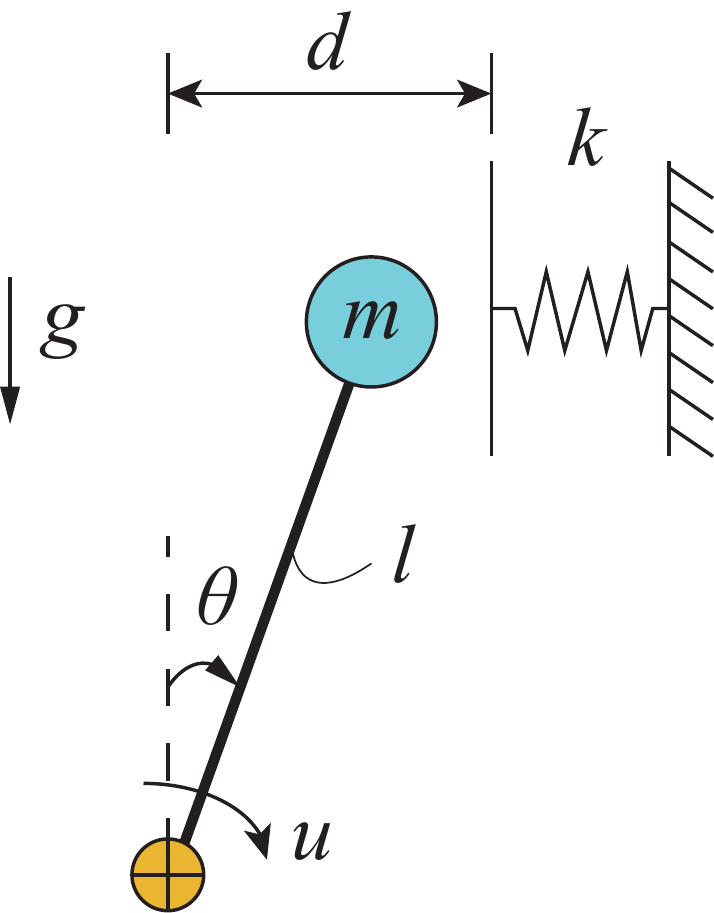}\\ 
  \small (A)
  \end{minipage}\ 
  \begin{minipage}{0.49\linewidth}
  \centering 
  \includegraphics[width=0.61\linewidth]{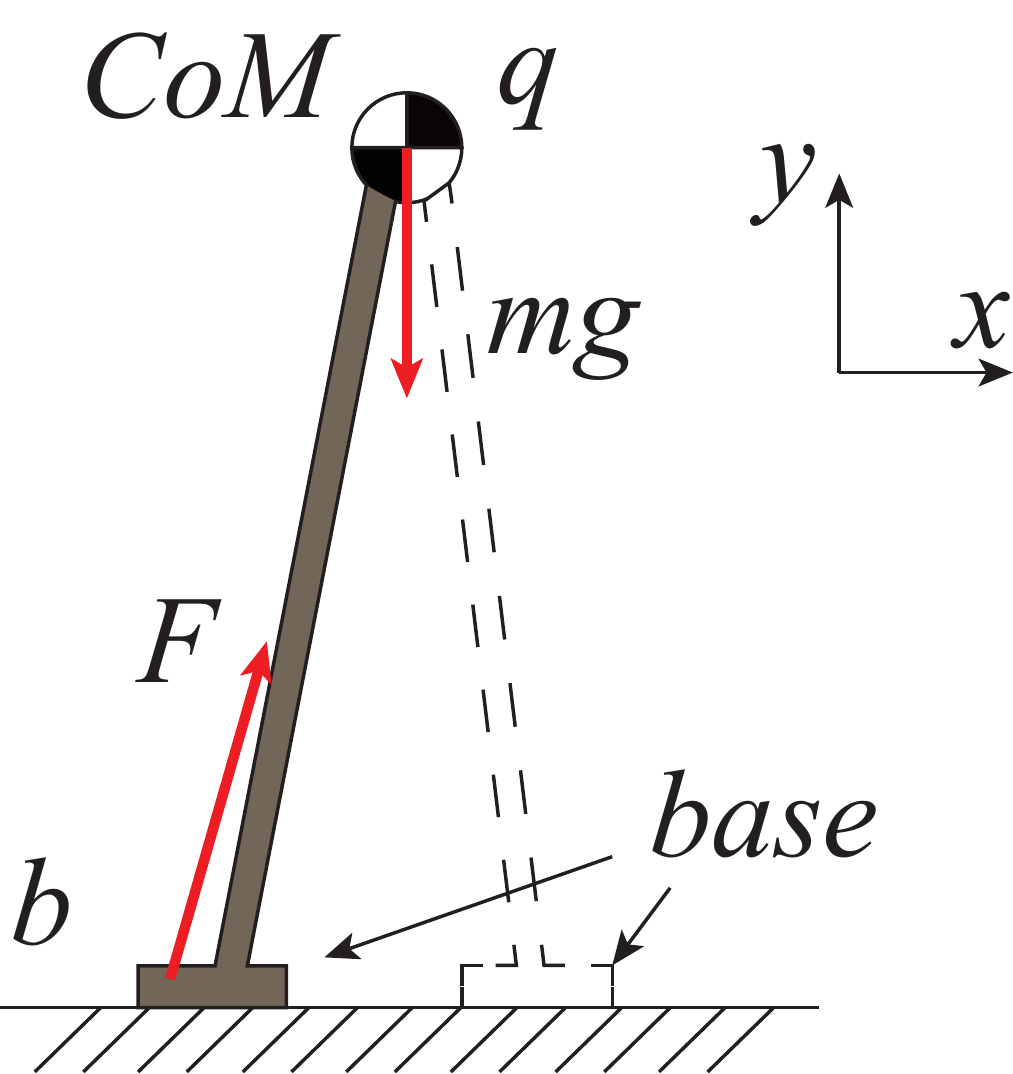} \\
  \small (B)
  \end{minipage}
  \caption{Examples used in this work. (A) Linear inverted pendulum with an elastic wall. (B) Variable height inverted pendulum (legged robot).}
  \label{fig:linear_inverted_pendulum_with_a_wall_pic}
\end{figure}

Piecewise-affine (PWA) systems, a special class of hybrid systems, where the state-input space is divided into polytopic regions with each region associated with a different affine dynamics equation, have been extensively studied.
PWA systems can arise by linearizing the dynamics of a nonlinear/hybrid system at a few points in the state space \cite{marcucci2017approximate,deits2018lvis}.
The optimal control policy for PWA systems can be computed offline by (i) dynamic programming and multi-parametric convex programming \cite{baoti2006constrained, baotic2003infinite, baotic2003new, borrelli2003efficient, baric2005optimal, christophersen2005optimal,  borrelli2005dynamic, baric2008efficient}, or (ii) representing the system as a mixed logical dynamical (MLD) system and then applying multi-parametric mixed integer convex programming \cite{bemporad2000optimal,bemporad2000piecewise,bemporad2002optimal},
or (iii) enumerating mode switch sequences and applying multi-parametric convex programming \cite{mayne2002optimal}.
In general, the computational time of these methods grows exponentially as the number of time steps grow until, if possible, convergence.
Lyapunov-function-based approaches have also been used for the feedback control of PWA systems, where linear or bilinear matrix inequalities are formed to search for a PWA controller and a piecewise-quadratic Lyapunov function that proves the stability of the closed-loop system \cite{rodrigues2002dynamic,lazar2006model,han2017feedback}.
The main drawbacks of these approaches are that the formulations are usually too conservative and hence may not be able to find feasible solutions for practical problems, and that bilinear matrix inequalities are non-convex and often intractible.
We think that piecewise polynomial functions, which include PWA functions as a special case, are better characterizations of the true dynamics of many systems.

Recent years has seen the development of the occupation measure approach \cite{lasserre2008nonlinear}.
The general framework of the approach is to first formulate the problem as an infinite-dimensional linear programming (LP) problem on measures and its dual on continuous functions, and to then approximate the LP by a hierarchy of finite-dimensional semidefinite programming (SDP) programs on moments of measures and their duals on sums-of-squares (SOS) polynomials.
The approach has been applied to approximating the region of attraction, the maximum controllable set, or the forward/backward reachable set for discrete-time/continuous-time autonomous/controlled hybrid/non-hybrid polynomial systems \cite{henrion2014convex, korda2013inner, korda2014controller, korda2014convex, shia2014convex, magron2017semidefinite}.
It has also been applied to controller synthesis for those systems \cite{korda2014controller,majumdar2014convex,savorgnan2009discrete,han2018controller,zhao2017control}.

The closest work to this paper is \cite{zhao2017control}, in which the authors used occupation measures to design controllers for continuous-time hybrid systems.
The main difference is that we focus on discrete-time systems as digital computers control robots via discrete signals.
Also, we usually test our closed-loop systems in simulators \cite{drake} that use time-stepping schemes and complementarity problems \cite{stewart2000implicit,chakraborty2007implicit} to model contacts.
Previous works \cite{korda2014controller,savorgnan2009discrete} on controller synthesis for discrete-time non-hybrid systems using occupation measures lack the notion of ``the state finally landing in the target set'', as is commonly used in continuous-time systems, and hence enforcing the system to reach a target set depends completely on the objective function, which can be hard to tune.
Our previous work \cite{han2018controller} solved this problem by proposing a new form of the Liouville equation.

In this paper, based on our previous work \cite{han2018controller}, we find what we call ``the one-step transition sets'' and formulate an optimization problem for discrete-time hybrid systems with the goal of steering as many initial states to the target set as possible in finite time.
As far as we know, this is the first time that occupation measure approach has been applied to controller synthesis for discrete-time hybrid polynomial systems.
The advantages of our approach include the following: 
(i) The approach only needs to solve SDPs, which are convex. For a fixed degree of SDP approximations, the computational complexity is polynomial in the state and control dimensions, and hence in principle our approach is more scalable than those based on dynamic programming.
(ii) The approach does not fix any mode switch sequences beforehand, and it finds them automatically.
(iii) The approach does controller synthesis for all initial states simultaneously instead of only one initial state.
(iv) The approach does not fix the number of time steps.
(v) As will be shown by the examples, the controller can take a simple form and is efficient to be applied online, implying low degrees of the SDP approximations work well in some cases.
In the last section, we will discuss the limitations of our approach. 
For example, the controllable set has to be checked a posteriori.

\subsection{Notations}

Let $\N$ (resp. $\R$) be the set of nonnegative integers (resp. real numbers).
Let $n\in \N$.
Let $\R[x]:=\R[x_1,\ldots,x_n]$ be the ring of polynomials in $x := (x_1,\ldots,x_n)$, and let $\R_{r}[x]$ be the set of polynomials in $\R[x]$ with degree at most $r$.
Let $\Sigma[x]$ denote the cone of SOS polynomials in $x$, and $\Sigma_r[x]$ the cone of SOS polynomials of degree up to $2r$.

Let $X \subseteq \R^n$ be a compact set.
Let $\mathcal{C}(X)$ denote the space of all continuous functions on $X$.
It is a Banach space equipped with the sup-norm.
Its topological dual $\mathcal{C}'(X)$ is the set of all continuous linear functionals on $\mathcal{C}(X)$.
$\mathcal{C}_+(X)$ denotes the cone of non-negative elements of $\mathcal{C}(X)$, i.e., it is the cone of non-negative functions on $X$.
If $\mu$ is a finite measure on the Borel $\sigma$-algebra $\mathcal{B}(X)$, then $\mu$ is a Radon measure \cite{royden1988real}.
Let $\mathcal{M}(X)$ denote the Banach space of finite signed Radon measures on $\mathcal{B}(X)$ equipped with the total variation norm, and $\mathcal{M}_+(X)$ is the cone of (unsigned) Radon measures.
By Riesz Representation Theorem, $\mathcal{M}(X)$ is isometrically isomorphic to  $\mathcal{C}'(X)$.
The topology in $\mathcal{M}_+(X)$ is the weak-star topology, while the topology in $\mathcal{C}_+(X)$ is the strong topology of uniform convergence.
For $\mu,\nu\in \mathcal{M}(X)$, if $\nu-\mu \in \mathcal{M}_+(X)$, then we say $\mu$ is dominated by $\nu$, and denote it by $\mu\leq \nu$.
For a measure space $(X,\mathcal{A},\mu)$ and $V \in \mathcal{A}$, let $\mathcal{A}|_V = \{A\subseteq V| A\in \mathcal{A} \}$ and let $\mu_V: \mathcal{A}|_V \to [0,+\infty]$ denote the restriction of $\mu$ on $\mathcal{A}|_V$, i.e., $\mu_V(A) = \mu(A)$, for $A \in \mathcal{A}|_V$.
More reviews on real and functional analysis can be found in \cite{folland2013real,conway2013course}.

\section{Problem formulation}
Let $n,m,s \in \N$.
Consider the discrete-time hybrid control-affine polynomial system $\mc{H} = (X,U,\mc{I},\mc{S},\phi)$:
\begin{itemize}
    \item $\mc{I} = \{1, \ldots , s\}$ is the set of indices of
the state space cells.
    Each (discrete state) $i\in\mc{I}$ is called a ``mode".
    \item The state constraint set $X \subseteq \R^n$ is the union of all state space cells $X_i$'s, $X = \bigcup_{i\in \mathcal{I}} X_i$.
    The state space cell $X_i \subseteq \R^n$ is a compact set with nonempty interior, $\forall i \in \mathcal{I}$.
    $X_i \cap X_j$ has empty interior, $\forall i\neq j \in \mathcal{I}$.
    \item $U \subseteq \R^m$ is the control input constraint set.
    \item The dynamics in the cell $X_i$ is given by
        $x_{t+1} = \phi_i(x_t,u_t) :=  f_i(x_t) + g_i(x_t) u_t, \text{ for } (x_t,u_t)\in X_i \times U$,
    where $f_i$ and $g_i$ are polynomials, $i\in\mc{I}$, and the vectors $x_t$ and $u_t$ represent the state and the control input at time $t$, respectively.
    The dynamics of the system at time $t$ is completely determined by the cell in which the state $x_t$ resides.
    The dynamics at the boundary $X_i \cap X_j$ is defined to be $\phi_k$, where $k = \min\{i,j\}$.
    \item $\mc{S} = \{(i,j)\in \mc{I}\times\mc{I}: \exists x_k \in X_i, u_k\in U, \text{ s.t. } x_{k+1} = \phi_i(x_k,u_k) \in X_{j}\}$ is the set of ordered pairs $(i,j)$ of indices denoting possible switches from cell $i$ to cell $j$.
\end{itemize}

\begin{problem}{1}
Given a discrete-time hybrid control-affine polynomial system $\mathcal{H}$ and a target set $Z \subseteq X$, design a state feedback controller $u: X \to U$ that maximizes the volume of the set $X_0$ where the trajectory starting from $X_0$ under $u$ reaches $Z$ in finite time.
\end{problem}

To solve this problem, we are going to design piecewise polynomial controllers expressed as $u = u^i(x) \in U$ if $x\in X_i$, where $u^i(x)$ is a polynomial in $x$, $i\in \mathcal{I}$.
In the end, our controller would be an approximation of the optimal controller, and we show by examples that the approximate controller is good enough for practical purposes.

Assume $X := \{ x\in \R^n| h^X_j(x) \geq 0, h^X_j(x)\in \R[x],j=1,\ldots,n_X\}$, 
and ${X_i} := \{ x\in \R^n| h^{X_i}_j(x) \geq 0, h^{X_i}_j(x)\in \R[x], j=1,\ldots,n_{X_i}\}, i \in \mc{I}$, are compact basic semi-algebraic sets with nonempty interior.
Furthermore, assume that the moments of the Lebesgue measure on $X_i$'s are available.
For example, $X_i$ can be an $n$-dimensional ball or box.
Assume that the origin is in the interior of $X_1$. 
This assumption excludes systems where the origin is at the boundaries of multiple cells.
By properly modifying the optimization formulation, we can deal with other classes of hybrid systems, but in this paper we only show how to work with this kind of hybrid systems.

Assume $U := \{ u\in \R^m| h^U_j(u) \geq 0, h^U_j(u)\in \R[u],j=1,\ldots,n_U\} = [a_1,b_1]\times \ldots \times [a_m,b_m]$.
In the sequel, without loss of generality, we assume  $U := [-1,1]^m$,
because the dynamics equations can be scaled and shifted arbitrarily.

Assume $Z := \{ x\in \R^n| h^Z_{j}(x)\geq 0, h^Z_j(x)\in \R[x],j=1,\ldots,n_Z\}$ is a compact basic semi-algebraic set with nonempty interior.
In practice, we may choose $Z$ to be a small ball or box around the origin so that after the system enters $Z$, we may turn on LQR or some other local control methods to regulate the system to the origin.
For this purpose, assume that the origin is in the interior of $Z$, and that $Z \subseteq X_1 \subseteq X$.

Define the one-step transition sets $Y_{ij}$, for any $(i,j) \in \mc{S}$, to be the preimage of $X_j$ under $f_i$ in $X_i \times U$, i.e.,
$Y_{ij} := \{(x,u) \in X_i \times U| f_i(x,u) \in X_j\}= f_i^{-1}(X_j) \bigcap X_i \times U
    = \{(x,u)\in X_i\times U| h_k^{X_j}(f_i(x,u)) \geq 0,k=1,\ldots,n_{X_j}\}$. 
The last equality implies that $Y_{ij}$ is a basic semi-algebraic set.
So we can write $Y_{ij} = \{(x,u)\in \R^{n+m}| h^{Y_{ij}}_k(x,u) \geq 0, h^{Y_{ij}}_k \in \R[x,u],
     k = 1,\ldots,n_{Y_{ij}}\}, \forall (i,j) \in \mc{S}$.
Since $X_j$ is compact and hence closed, and since $f_i$ is continuous, $f_i^{-1}(X_j)$ is closed.
Since $X_i \times U$ is compact, $Y_{ij}$, being a closed subset of a compact set, is compact.

Let $r^{X_{i}}_{j} := \lceil {\deg h^{X_i}_j / 2} \rceil, i\in \mathcal{I}, j=0,\ldots,n_{X_i}$.
Let $\mathbf{Q}^{X_i}_r := \big\{\sum_{j=0}^{n_{X_i}} \sigma_j(x) h^{X_i}_j(x): \sigma_j \in \Sigma_{r-r^{X_i}_j}[x], j=0,\ldots,n_{X_i} \big\}$ denote the $r$-truncated quadratic module generated by the defining polynomials of $X_i$, assuming $h^{X_i}_0(x) = 1$.
Analogously, we define $\mathbf{Q}^{X_i U}_r$, $\mathbf{Q}^Z_r$, and $\mathbf{Q}^{Y_{ij}}_r$.
Note that $\mathbf{Q}^{X_i U}_r$ is generated by defining polynomials of $X_i$ and $U$.

(Putinar's condition) A compact basic semialgebraic set defined by 
$\Omega := \{x \in \R^n \vert  h_j^\Omega(x) \geq 0, j = 1,\ldots,n_{\Omega}\}$ satisfies Putinar's condition if there exists $h \in \R[x]$ such that $h = \sigma_0 + \sum_{j=1}^{n_{\Omega}} \sigma_j h_j$ for some $\{\sigma_j\}_{j=0}^{n_{\Omega}} \subset \Sigma[x]$ and the level set $\{x\in \R^n\vert h(x) \geq 0\}$ is compact.

Putinar's condition can be satisfied by including $N - ||x||_2^2$ in the defining polynomials of $\Omega$ for some large $N \in \R$.
If $\Omega$ satisfies Putinar's condition, then  Putinar's Positivstellensatz \cite{putinar1993positive} says that any positive polynomial on $\Omega$ has an SOS representation.

\section{Optimization formulation}
\subsection{Discrete-time controlled Liouville equation}
The Liouville equation describes the evolution of the state distribution of the system over time.
For continuous-time systems, it takes the form of partial differential equations (see e.g. \cite{henrion2014convex}).
For discrete-time systems, it is realized by the push forward operator \cite{magron2017semidefinite, han2018controller}. 
In the following, we introduce the discrete-time controlled Liouville equation in the non-hybrid setting as proposed in \cite{han2018controller}, and in the next subsection we shall see how it enables us to formulate optimization problems for hybrid systems.

Given measurable spaces $(S_1,\mc{A}_1)$ and $(S_2,\mc{A}_2)$, a measurable function $f: S_1 \to S_2$, and a measure $\nu: \mc{A}_1 \to [0, +\infty]$, the pushforward measure $f_*\nu:\mc{A}_2 \to [0,+\infty]$ is defined to be $f_* \nu(A) := \nu(f^{-1}(A))$, for all $A\in \mc{A}_2$.

Assume $s=1$, i.e. $X = X_1$ (just for the rest of this subsection).
Let $X_0, X_T \subseteq X$ be the measurable sets of all possible initial states and all possible final states of system trajectories, respectively.
Define $\pi$ to be the projection $\pi: X\times U \to X, (x,u) \mapsto x$.
It simply extracts the state from a state-input pair.
$\phi: X\times U \to X$ describes the system dynamics as defined in the previous section.
The discrete-time controlled Liouville equation is 
\begin{align}
        \mu + \pi_* \nu = \phi_* \nu + \mu_0, \label{Liouville}
\end{align}
where $\mu_0 \in \mathcal{M}_+(X_0),\mu\in \mathcal{M}_+(X_T)$ and $\nu \in \mathcal{M}_+(X\times U)$.

The initial measure $\mu_0$ can be viewed as the distribution of the mass of the initial states of the system trajectories (not necessarily normalized to 1).
The occupation measure $\nu$ describes the volume occupied by the trajectories.
The final measure $\mu$ can be viewed as the distribution of the mass of the final states of the system trajectories. 
The measures $\mu_0,\nu$, and $\mu$ will be decision variables in our optimization, which enables us to search over trajectories with certain properties.

\subsection{Primal-dual infinite-dimensional LP}

Define the projections $\pi_i : X_i \times U \to X_i, (x,u) \mapsto x, i\in\mc{I}$.
The infinite-dimensional LP on measures is formulated as follows:
\begin{align}
& p := \sup_{\mu^i_0,\hat{\mu}^i_0, \mu, \nu_i,\mu_{ij}}\   \sum_{i\in \mc{I}} \int_{X} 1 d\mu^i_0 \  \text{ subject to:} \label{lp_primal}\\
&\displaystyle \sum_{j:(i,j)\in \mc{S}} \pi_{i*}\mu_{ij} + \pi_{i*} \nu_i = \phi_{i*} \nu_i + \mu^i_0 + \sum_{j:(j,i)\in \mc{S}} \phi_{j*} \mu_{ji},\nonumber\\
& \quad\quad\quad\quad\quad\quad\quad\quad\quad\quad\quad\quad\quad\quad\quad\quad\quad \text{ for } i\neq 1, i \in \mathcal{I},\nonumber\\
&\mu + \pi_{i*}\nu_i = \phi_{i*}\nu_i + \mu^i_0 + \sum_{j:(j,i)\in \mc{S}} \phi_{j*} \mu_{ji}, \text{ for } i = 1,\nonumber\\
&\mu^i_0 + \hat{\mu}^i_0 = \lambda_{X_i}, \mu^i_0,\hat{\mu}^i_0 \in \mathcal{M}_+(X_i),\forall i\in\mc{I},\nonumber\\
&\mu \in  \mathcal{M}_+(Z),\nu_i \in \mathcal{M}_+(X_i\times U),\forall i\in\mc{I},\nonumber\\
&\mu_{ij} \in \mc{M}_+(Y_{ij}), \forall (i,j)\in \mc{S}.\nonumber
\end{align}

Remember that we assume $Z\subseteq X_1$.
The first constraint is a direct application of the Liouville equation.
In this constraint, we are considering any mode $i$ whose state space cell does not contain the origin, i.e., $1\neq i\in \mc{I}$.
The measure $\nu_i$ is the occupation measure describing the volume occupied by system trajectories in $X_i$.
The final measure $\sum_{j:(i,j)\in \mc{S}} \pi_{i*}\mu_{ij}$ consists of the measures supported on the one-step transition sets $Y_{ij}$ for all $j$ such that $(i,j) \in \mc{S}$, meaning that if $f_i$ is applied again to those final states, the system will leave $X_i$ and enter other modes.
The initial measure $\mu^i_0 + \sum_{j:(j,i)\in \mc{S}} \phi_{j*} \mu_{ji}$ consists of two parts, $\mu^i_0$ describing the distribution of initial states originated from $X_i$, and $\sum_{j:(j,i)\in \mc{S}} \phi_{j*} \mu_{ji}$ describing the distribution of states coming from other modes.
The second constraint is the Liouville equation for mode 1.
It differs from the first constraint in the final measure, which is a measure $\mu$ supported on the target set $Z$.
This is crucial to our goal of controlling the system to the target set, because
adding the term $\sum_{j:(i,j)\in \mc{S}} \pi_{i*}\mu_{ij}$ to the final measure in the second constraint would cause chattering effects -- for many initial states, the system would bounce between two modes and finally stop somewhere at the boundary of two modes.
The third constraint ensures that any initial measure originated from mode $i$ is dominated by the Lebesgue measure on $X_i$, $\forall i$.
The objective is to maximize the sum of the mass of initial measures originated from all modes.
The combination of the third constraint and the objective means maximizing the volume of the set of initial states that can be controlled to the target set.
The whole optimization design is specific to solving Problem 1.
For other optimal control purposes, the objective can be modified to be some other reward or cost functions, and the constraints can be modified accordingly.

The dual LP on continuous functions is given by 
\begin{align}
    & \inf_{w_i,v_i,i\in\mc{I}}  \ \sum_{i\in\mc{I}}\int_{X_i} w_i(x) d\lambda_{X_i} \  \text{ subject to:} \label{lp_dual}\\
    & \ v_i(x) - v_i(\phi_i(x,u)) \geq 0, \forall x \in X_i, i\in\mc{I},\forall u\in U,\nonumber\\
    & \ v_i(x) - v_j(\phi_i(x,u)) \geq 0, \forall (x,u) \in Y_{ij}, i \neq 1, (i,j)\in\mc{S},\nonumber\\
    & \ \quad \quad \  - v_j(\phi_i(x,u)) \geq 0, \forall (x,u) \in Y_{ij}, i = 1,(i,j)\in\mc{S},\nonumber\\
    & \ w_i(x) - v_i(x) - 1 \geq 0, w_i(x) \geq 0,\forall x \in X_i,i\in\mc{I}, \nonumber\\
    & \ v_1(x) \geq 0, \forall x \in Z, \ v_i,w_i \in \mathcal{C}(X_i),i\in\mc{I}.\nonumber
\end{align}

\section{Semidefinite relaxations}
Although we cannot directly solve the infinite-dimensional LP's, we can approximate them by finite-dimensional SDP's and extract the controller from the solution to the SDP's.
Semidefinite approximation to the infinite-dimensional LP is quite standard in the literature (see e.g. \cite{lasserre2008nonlinear}). 
It is based on the idea that a measure can be characterized by its sequence of moments, just as a signal can be characterized by its sequence of Fourier coefficients.
In this section, we first introduce some background knowledge about moments of measures, which can also be found in \cite{lasserre2010moments}.
Next we formulate the relaxed SDP's on moments of measures and their duals on SOS polynomials.
Finally, we briefly mention how to extract controllers. 

\subsection{Preliminaries} 
Any polynomial $p(x)\in \R[x]$ can be expressed in the monomial basis as $p(x) = \sum_{\alpha} p_\alpha x^{\alpha}$, where $\alpha \in \N^n$, and $p(x)$ can be identified with its vector of coefficients $p:=(p_\alpha)$ indexed by $\alpha$.
The (sequence of) moments $y:=(y_\alpha)$ for a measure $\mu$ is defined to be 
$y_\alpha := \int x^\alpha d\mu, \alpha \in \N^n$.
Given a sequence of real numbers $y=(y_\alpha)$, $y$ is not necessarily a sequence of moments for some measure $\mu$.
If it is, then $\mu$ is called a representing measure for $y$.
Given any $y$, we define the linear functional $\ell_y: \R[x] \to \R$ by
$\ell_y(p(x)):= p^\top y = \sum_\alpha p_\alpha y_\alpha$.
Integration of a polynomial $p$ against a measure $\mu$ with the sequence of moments $y$ can be expressed as a linear functional $\ell_y$: $\int p d\mu = \int \sum_\alpha p_\alpha x^\alpha d\mu = \sum_\alpha p_\alpha \int x^\alpha d\mu = \sum_\alpha  p_\alpha y_\alpha$.

Given $r\in \N$, define $\N^n_r = \{\beta\in \N^n: |\beta|:=\sum_i \beta_i \leq r\}$.
Define the moment matrix $M_r(y)$ of order $r$ with entries indexed by multi-indices $\alpha$ (rows) and $\beta$ (columns) to be $[M_r(y)]_{\alpha,\beta} := \ell_y(x^\alpha x^\beta) = y_{\alpha+\beta}, \forall \alpha,\beta \in \N^n_r$.
The moment matrix can be expressed as a bilinear form $\left< \cdot , \cdot \right>_y$ on $\R[x]_r$:
$\left< p, q \right>_y := \ell_y(pq) = p^\top M_r(y) q, \forall p,q \in \R[x]_r$.
If $y$ has a representing measure, then $M_r(y) \succeq 0$, $\forall r\in \N$.
However, the converse is generally not true.

Given a polynomial $u(x)\in \R[x]$ with coefficient vector $u = (u_\gamma)$, define the localizing matrix w.r.t. $y$ and $u$ to be the matrix indexed by multi-indices $\alpha$ (rows) and $\beta$ (columns)
$[M_r(uy)]_{\alpha,\beta} := \ell_y(u(x)x^\alpha x^\beta) = \sum_{\gamma} u_{\gamma}y_{\gamma + \alpha + \beta}, \forall \alpha,\beta \in \N^n_r$.
Similar to the moment matrix, we have 
$\left< p,M_r(uy)q \right> = \ell_y(upq) = p^\top M_r(uy)q, \forall p,q \in \R[x]_r$.
If $y$ has a representing measure $\mu$, then $M_r(uy)\succeq 0$ whenever the support of $\mu$ is contained in $\{x\in \R^n: u(x) \geq 0\}$.
Conversely, if $\Omega$ is a compact basic semi-algebraic set as defined in Section II, if Putinar's condition holds, and if $M_r(h^\Omega_j y) \succeq 0, j=0,\ldots,n_\Omega, \forall r$, then $y$ has a finite Borel representing measure with support contained in $\Omega$ (Theorem 3.8(b) in \cite{lasserre2010moments}).

\subsection{Primal-dual finite-dimensional SDP}
For each $r \geq r_{min} := \max_{k_1, k_2, k_3, k_4} \{r^{X_i}_{k_1} (i\in\mc{I}), r^U_{k_2},\\ r^{Y_{ij}}_{k_3} ((i,j) \in \mc{S}), r^Z_{k_4} \}$, let $y^i_0 = (y^i_{0,\beta}),\beta\in \N^n_{2r}$, be the finite sequence of moments up to degree $2r$ of the measure $\mu^i_0$.
Similarly, $y_1, \hat{y}^i_0, y^{X_i}$, $z_i$, $y_{ij}$ are finite sequences of moments up to degree $2r$ associated with measures $\mu,\hat{\mu}^i_0$, $\lambda_{X_i}$, $\nu_i$, and $\mu_{ij}$, respectively.
Let $d_i := $ degree  $\phi_i$.
The infinite-dimensional LP on measures (\ref{lp_primal}) can be relaxed with the following semidefinite program on moments of measures:
\begin{align}
    & p_r := \sup_{y_0^i,y_1, \hat{y}^i_0, z_i, y_{ij}}\  \sum_{i\in\mc{I}} y^i_{0,0} \  \text{ subject to:}  \label{sdp_primal}\\
    &\ \sum_{j:(i,j)\in\mc{S}} \ell_{y_{ij}}(x^\beta) + \ell_{z_i}(x^\beta) = \ell_{z_i}(\phi_i(x,u)^\beta) + y^i_{0,\beta} \nonumber\\
    &\quad\quad\ + \sum_{j:(j,i)\in\mc{S}} \ell_{y_{ji}}(\phi_j(x,u)^\beta),\forall \beta\in\N^n_{2r},1\neq i\in\mc{I}, \nonumber\\
    &\ y_{1,\beta} + \ell_{z_i}(x^\beta) = \ell_{z_i}(\phi_i(x,u)^\beta) + y^i_{0,\beta}\nonumber\\
    &\quad\quad\ + \sum_{j:(j,i)\in\mc{S}} \ell_{y_{ji}}(\phi_j(x,u)^\beta), 
    \forall \beta \in \N^n_{2r}, i=1,\nonumber\\
    & \ y^i_{0,\beta} + \hat{y}^i_{0,\beta} = y^{X_i}_\beta, \forall \beta \in \N^n_{2r},i\in\mc{I},\nonumber\\
    & \ \mathbf{M}_{r-r^{X_i}_k}(h^{X_i}_k y^i_0) \succeq 0, k = 1,\ldots,n_{X_i},i\in\mc{I},\nonumber\\
    & \ \mathbf{M}_{r-r^{X_i}_k}(h^{X_i}_k \hat{y}^i_0) \succeq 0,k = 1,\ldots,n_{X_i},i\in\mc{I},\nonumber\\
    & \ \mathbf{M}_{rd_i-r^{X_i}_k}(h^{X_i}_k z_i) \succeq 0,k = 1,\ldots,n_{X_i},i\in\mc{I},\nonumber\\
    & \ \mathbf{M}_{rd_i-r^U_k}(h^U_k z_i) \succeq 0,k = 1,\ldots,n_U,i\in\mc{I},\nonumber\\
    & \ \mathbf{M}_{rd_i-r^{Y_{ij}}_k}(h^{Y_{ij}}_k y_{ij}) \succeq 0, k = 1,\ldots,n_{Y_{ij}},(i,j) \in \mc{S}, \nonumber\\
    & \ \mathbf{M}_{r-r^Z_k}(h^Z_k y_1) \succeq 0,k = 1,\ldots,n_Z. \nonumber
\end{align}
\begin{proposition}{1}
Assume all assumptions in Section II are satisfied.
Assume the sets $X_i$ for $i\in \mathcal{I}$, $Y_{ij}$ for $(i,j)\in \mathcal{S}$, and $Z$ all satisfy Putinar's  condition.
Suppose there exists $r_0 \geq r_{min}$ such that (i) for every $r \geq r_0, r\in \N$, SDP (\ref{sdp_primal}) is feasible and the optimal solution to SDP (\ref{sdp_primal}) exists, and (ii) there exists a ball of constant radius (independent of $r$) in the solution space such that it contains the feasible set of the SDP (\ref{sdp_primal}) for all $r \geq r_0, r \in \N$. 
Then $\lim_{r\to \infty} p_r = p$.
\end{proposition}

The dual of (\ref{sdp_primal}) is the following SDP on polynomials of degrees up to $2r$:
\begin{align}
    &\inf_{v,w}  \ \sum_{i\in\mc{I}} \sum_{\beta\in\N^n_{2r}} (w_i)_\beta y_\beta^{X_i} \ \text{ subject to:} \label{sdp_dual}\\
    & \ v_i -  v_i\circ \phi_i   \in \mathbf{Q}^{X_iU}_{rd_i}, i\in\mc{I}, \nonumber\\
    & \ v_i -v_j\circ \phi_i \in \mathbf{Q}^{Y_{ij}}_{rd_i},i\neq 1,(i,j)\in \mc{S},\nonumber\\
    & \ \ \ -v_j\circ \phi_i \in \mathbf{Q}^{Y_{ij}}_{rd_i},i= 1,(i,j)\in \mc{S},\nonumber\\
    & \ w_i - v_i - 1\in \mathbf{Q}^{X_i}_{r}, w_i \in \mathbf{Q}^{X_i}_{r}, i\in\mc{I}, \nonumber\\
    & \ v_1 \in \mathbf{Q}^Z_{r}, v_i,w_i \in \R_{2r}[x],i\in\mc{I}. \nonumber
\end{align}

The dual SDP (\ref{sdp_dual}) is a strengthening of the dual LP (\ref{lp_dual}) by requiring nonnegative polynomials in (\ref{lp_dual}) to be SOS polynomials up to certain degrees.
In practice, we use off-the-shelf numerical solvers to solve the dual SDP (\ref{sdp_dual}) and simultaneously get information about the primal SDP (\ref{sdp_primal}).
Usually $r$ has to be less than 10 due to limitations on current SDP solvers, but such $r$ already provides good controllers for many practical problems.
For fixed $r$ or for $r$ less than a fixed constant, the computational complexity is polynomial in the state and control input dimensions.
 

\subsection{Controller extraction}
We extract a polynomial controller for each mode. 
Combining these controllers, we have a piecewise polynomial controller on the entire state space.
The procedure of extracting a controller for mode $i$ from the moments of the measure $\nu_i$ is the same as in Section IV.C. of \cite{han2018controller}.

However, the extracted controller may not always satisfy the control input constraints.
The easiest remedy is to limit the control input to be the boundary values, $\pm 1$, if the constraints are violated.
For both examples in the next section, we used this method, and we noticed that most of the time, the control input constraints were not violated.
Another method is to solve an SOS optimization problem as in \cite{korda2014controller}.

\section{Examples}
We illustrate our controller synthesis method on some discrete-time hybrid polynomial systems.
All computations are done using MATLAB 2016b, the SDP solver MOSEK 8, and the polynomial optimization toolbox Spotless \cite{tobenkin2013spotless}.

\subsection{Linear inverted pendulum with a wall}

Consider the problem of balancing a linear inverted pendulum to its upright position with the existence of a nearby elastic wall, as depicted in Figure \ref{fig:linear_inverted_pendulum_with_a_wall_pic}(A) (Example A in \cite{marcucci2017approximate}).
The system can be modelled as a hybrid system with two modes: pendulum not in contact with the wall (mode 1), and pendulum in contact with the wall (mode 2).
In this example, we linearize the dynamics and work with the PWA system as in \cite{marcucci2017approximate}.
(We could have approximated the system with higher order polynomials, synthesized the controller, and run on the real system as in Example E in \cite{han2018controller}, but we are more interested in the comparison with the traditional model predictive control (MPC) approach.)

Let $m = 1, l = 1, d = 0.1, g = 10$, and $k = 1000$.
Let the state be $x = (\theta,\dot{\theta})$.
Linearizing the dynamics around the upright position $\theta = 0$, and discretizing the model with the explicit Euler scheme with the sampling time $\delta t = 0.01$, the system becomes a piecewise-affine system $x^+ = A_i x + B_i u + a_i, x \in X_i, i\in \mc{I} = \{1,2\}$, where the dynamics and the state space are the same as given in Example A of \cite{marcucci2017approximate}.
The target set is
$Z = \{ x\in \R^2| |x_1| \leq 0.03, |x_2|\leq 0.1\}$, which is inside the maximum LQR-control invariant set.

\begin{figure*}
  \includegraphics[width=0.32\linewidth]{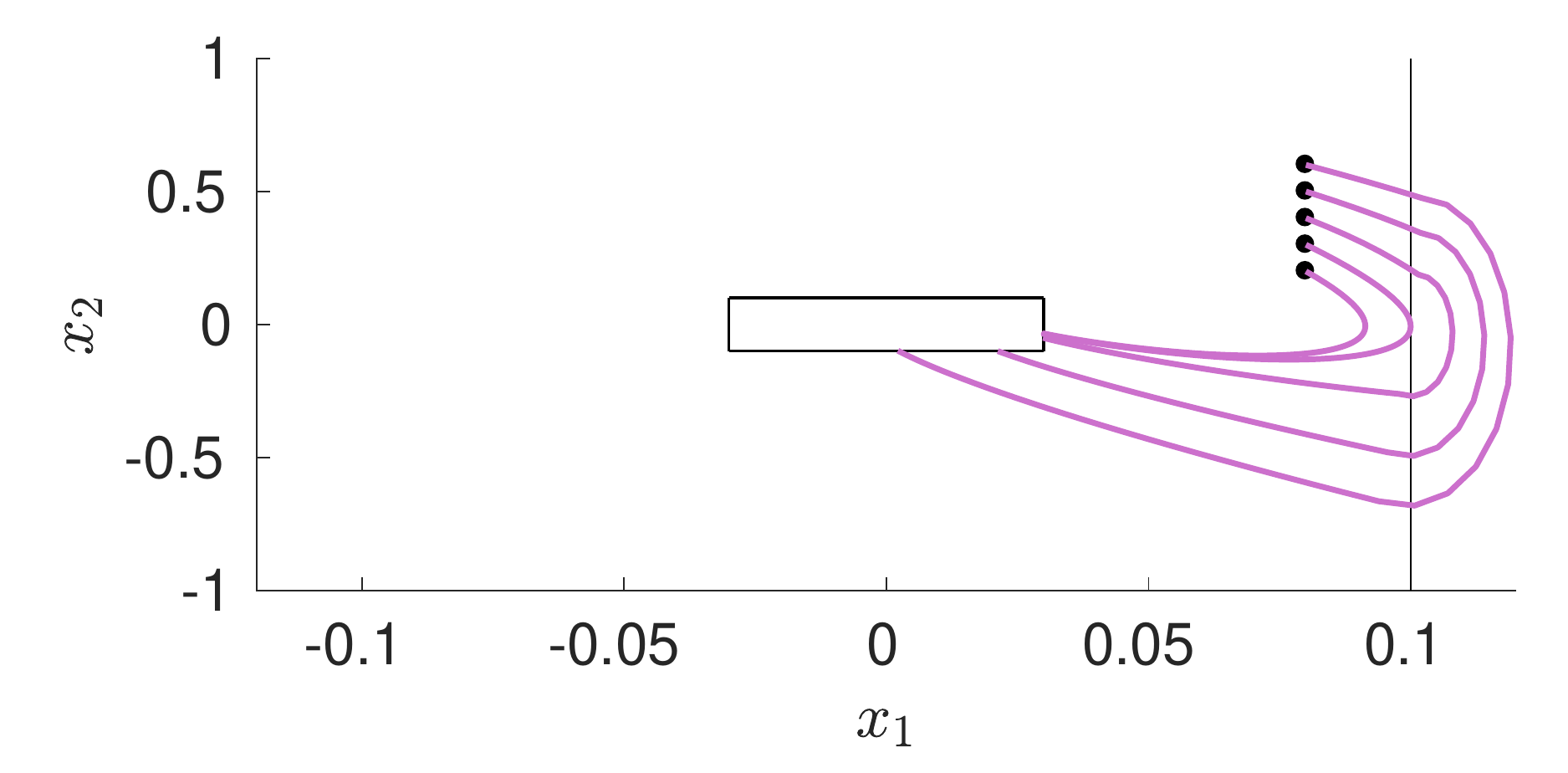}
  \includegraphics[width=0.32\linewidth]{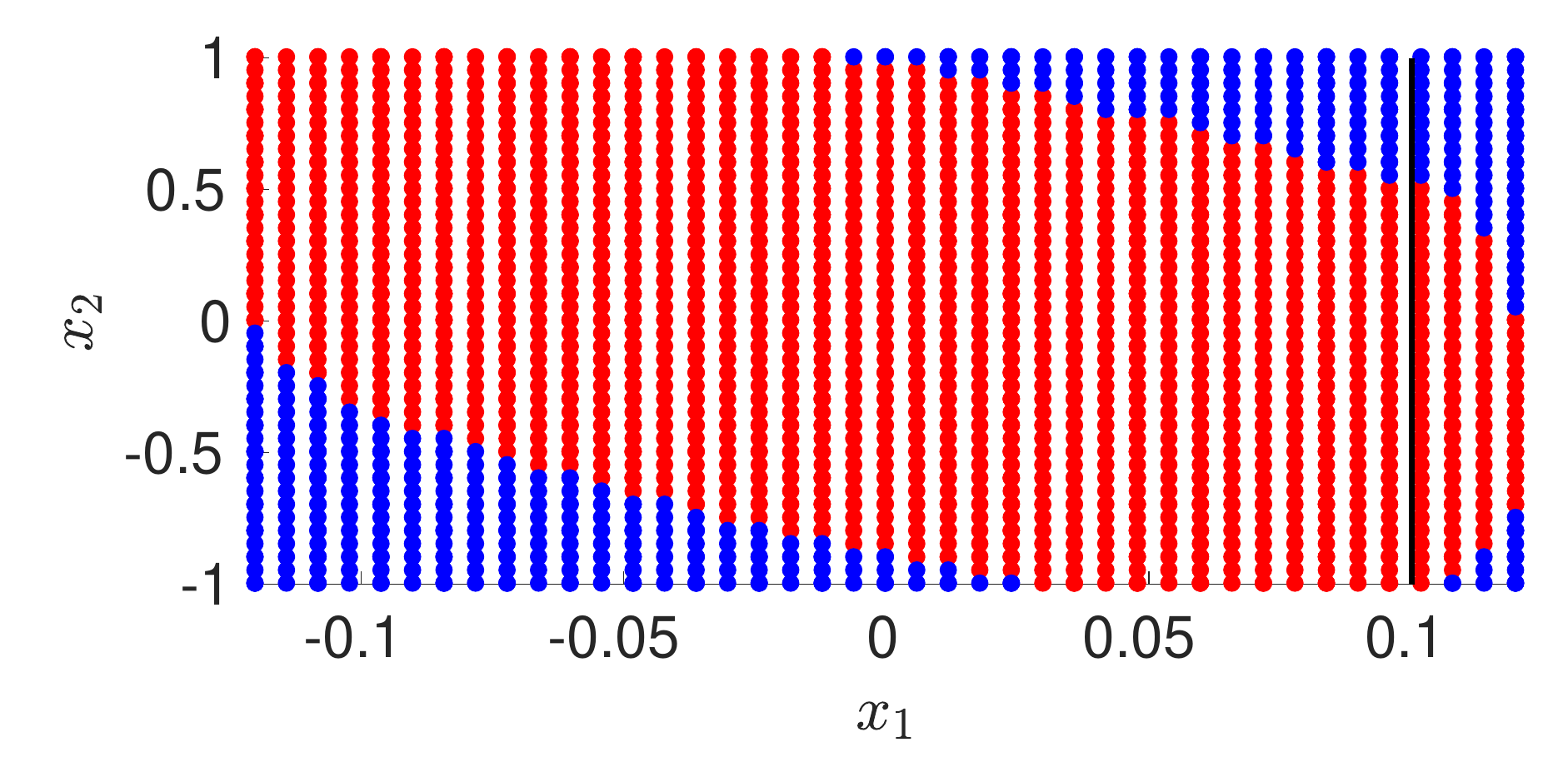}
  \includegraphics[width=.3\linewidth]{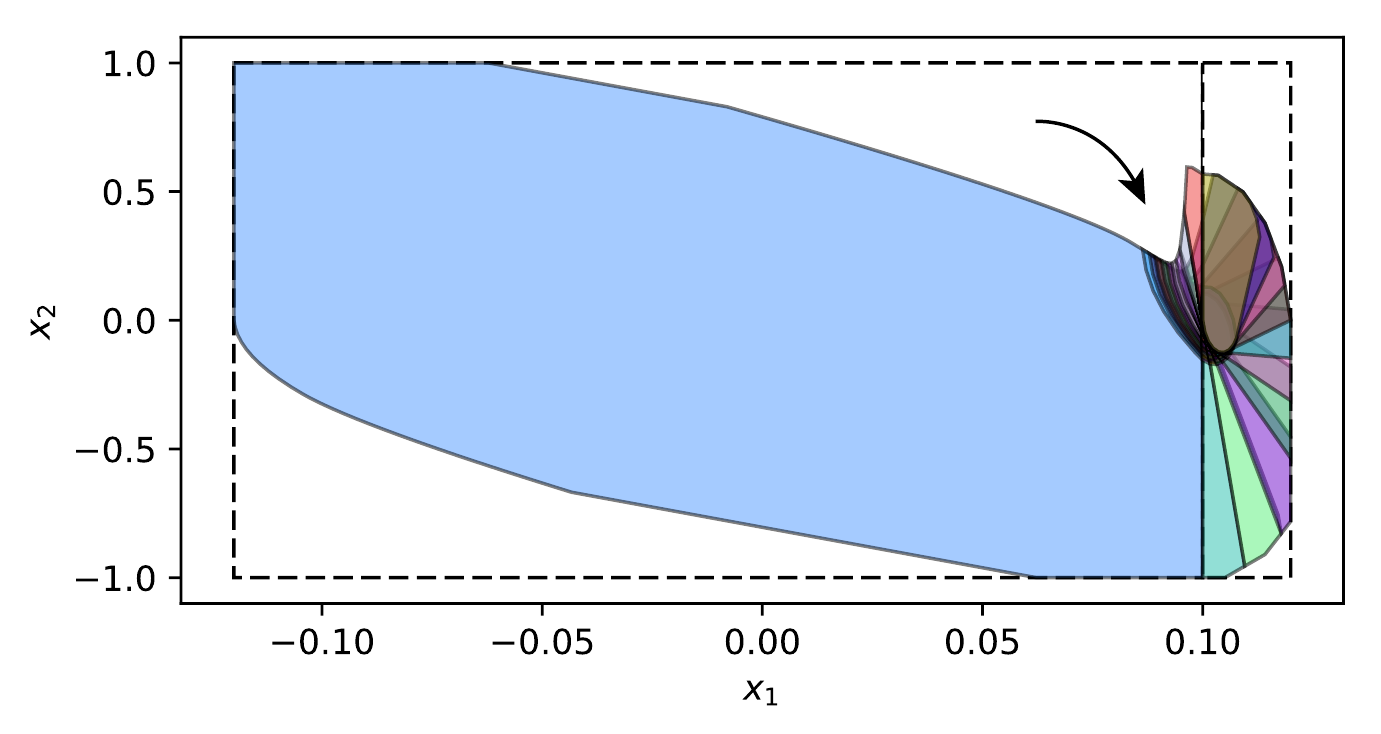}
  \caption{Linear inverted pendulum with a wall. Left: trajectories of five initial states under the occupation-measure-derived controller. Middle: controllable states under the occupation-measure-derived controller plotted in red and uncontrollable states plotted in blue. Right: feasible sets of the MPC approach.}
  \label{fig:inverted_pendulum_with_a_wall}
\end{figure*}

We search for a piecewise-affine controller, and the extracted controller is 
\begin{align*}
    & u^{1}(x) = 0.10336 - 6.7202 x_1 - 1.6978 x_2, \text{ for } x\in X_1,\\
    & u^{2}(x) = -0.62962 + 5.4774 x_1 - 0.60315 x_2, \text{ for } x\in X_2.
\end{align*}
The first two plots in Figure \ref{fig:inverted_pendulum_with_a_wall} show trajectories of five initial states $(0.08,0.2+i 0.1)$ for $i = 0,1,\ldots,4$, under the extracted controller and the uniformly sampled initial states with controllable states in red and uncontrollable states in blue.
The mode switch sequences are not specified beforehand, and it turns out that small differences in the initial states can result in very different mode switch sequences as shown in the trajectory plot.

We compare our method with the standard MPC approach.
With the MPC approach, we use a binary variable for each time step to indicate whether the system mode at that time step is mode 1 or mode 2.
Therefore, the MPC approach amounts to solving online a mixed integer quadratic programming (MIQP) problem with the terminal set being the maximum LQR-control invariant set and the terminal function being the solution to the Riccati equation.
Fixing a mode sequence over the time horizon, the MIQP problem becomes a QP problem, and the feasible set of the QP is a polytope.
The union of the feasible sets of the QP problems over all possible mode sequences is the feasible set of the MIPQ problem.
In this case (with the terminal set and the terminal cost as described above), the feasible set is the same as the controllable set.
The feasible set of the MIQP problem with time horizon $T = 10$ is depicted in the right plot (taken from Figure 5 of \cite{marcucci2017approximate}) of Figure \ref{fig:inverted_pendulum_with_a_wall}.
We choose the time horizon to be 10, because on an Intel i5, 2.3 GHz machine using Gurobi 8.0.0, the worst computation time for solving an MIQP with the time horizon of 10 already exceeds the sampling time 0.01 s. 
Each polytope in the right plot represents a feasible set of a QP problem with a certain mode sequence.
For example, the large blue region on the left represents the feasible set of the QP problem with the mode sequence being all 1's for the 10 time steps.

Our approach has three advantages.
First, we have a simpler controller.
Our controller is piecewise-affine with only two pieces, while the MPC controller is much more complicated.
Second, we have a larger controllable set (noticing our controllable set includes the white region pointed by an arrow in the right plot of Figure \ref{fig:inverted_pendulum_with_a_wall}).
This is because the MPC approach, restricted by the sampling time, has a limited time horizon, while our approach computes solutions for all finite time horizons.
Third, the online computation time of our approach is negligible, while the MPC approach needs to solve an MIQP problem online, which is $> 0.01$s in the worst case for a time horizon of 10.
It is amazing that our approach produces such a simple controller that controls such a large region.

\subsection{Variable height inverted pendulum (legged robot)}

\begin{figure}
    \centering
    \includegraphics[width=0.49\linewidth]{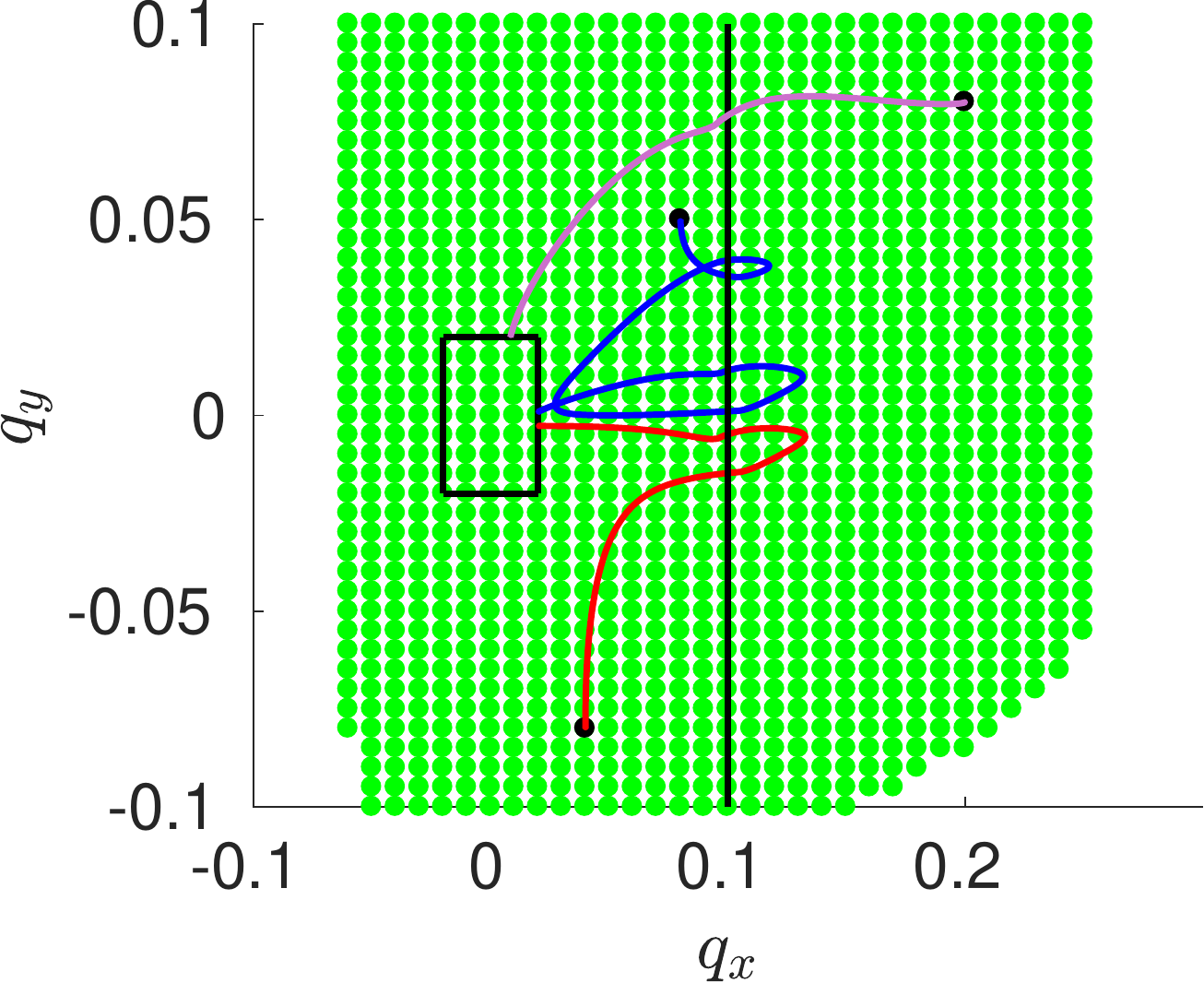}
    \includegraphics[width=0.49\linewidth]{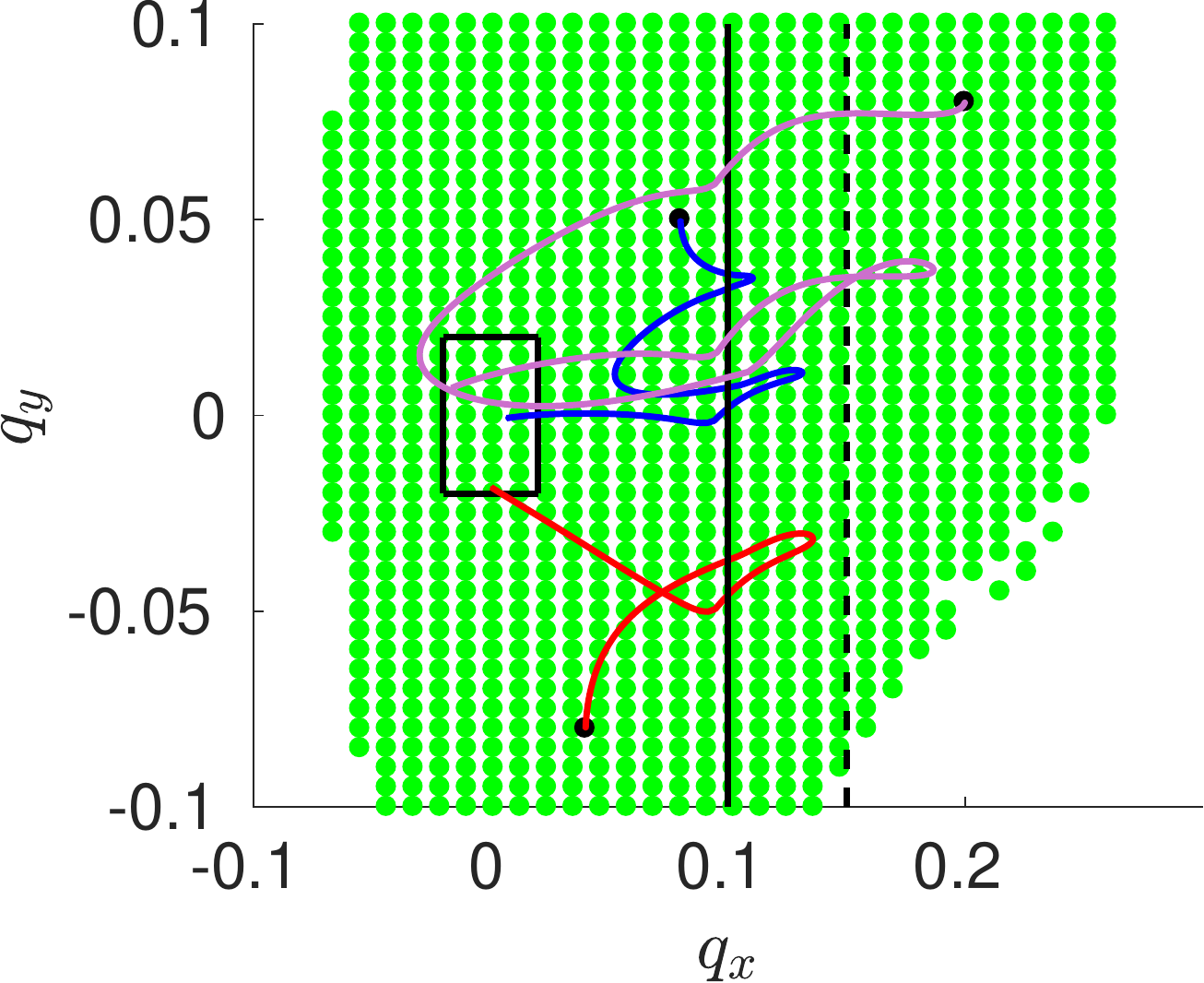}
    \caption{The $q_x-q_y$ sections of the state space, where other states are 0. Controllable points are plotted in green. The black box in each plot indicates the target set $Z$. The vertical solid line is the boundary of two modes. The vertical dashed line indicates a boundary of a base. $q_y$ is the deviation from the desired height ($1$ m above the ground). Three curves in each plot are the projections of the trajectories of three initial states $x_0 = [q_x,q_y,\dot{q}_x,\dot{q}_y] = [0.2,0.08,0,0],[0.08,0.05,0,0]$, and $[0.04,-0.08,0,0]$ onto the plane.}
    \label{fig:variable_height_inverted_pendulum_controllable_points}
\end{figure}

Consider balancing a legged robot modelled as a variable height inverted pendulum \cite{koolen2016balance,posa2017balancing}, as depicted in Figure \ref{fig:linear_inverted_pendulum_with_a_wall_pic}(B).
In this example, we assume there is one massless base (foot), and the robot can place it on two fixed places.
The center of pressure $b$ can range from anywhere in the base.
The center of mass (CoM) $q$ is constrained in a box area, independent of the distance to the center of base (CoB), the middle point of base.
The force exerted by the ground is $F = m(q-b)u$, where $u$ controls the magnitude of the force.
The dynamics of the system is given by 
    $m\ddot{q} = -mg + m(q-b)u$.
Both $b$ and $u$ are control inputs, and $b$ is hybrid with two possible steps to take on.
The system is a hybrid polynomial system with 2 modes, 4 states, and 2 control inputs.
The goal is to balance the CoM to the upright position, which is 1 m above the CoB.
Assume the length of the base is $0.2$ m.
We search for two 3rd order piecewise polynomial controllers $b$ and $u$.
The computation time on an Intel i7 3.3 GHz, 32 GB RAM machine was about 10 minutes.
Figure \ref{fig:variable_height_inverted_pendulum_controllable_points} shows the controllable points in the 2-dimensional $q_x-q_y$ plane sections of the 4-dimensional state space, where $q_x$ and $q_y$ are the horizontal and vertical displacement of the CoM from the desired position.
Projections of three trajectories onto the $q_x-q_y$ plane are also plotted.
In the first experiment (left), the base can be placed on the intervals $[-0.1,0.1]$ and $[0.1,0.3]$, i.e., for mode 1, $b \in [-0.1,0.1]$, and for mode 2, $b \in [0.1,0.3]$. 
In the second experiment (right), the base can be placed on the intervals $[-0.1,0.1]$ and $[0.15,0.35]$.
In both experiments, $q_x = 0.1$ is the boundary of two modes, i.e., if $q_x<0.1$, then $b\in [-0.1,0.1]$, otherwise $b$ is in the other interval ($[0.1,0.3]$ or $[0.15,0.35]$).

\section{Conclusion and discussion}
We have presented a controller synthesis method for discrete-time hybrid polynomial systems via the notion of occupation measures.
We noticed that controllers of certain degrees work better for certain systems.
For example, for PWA systems we found PWA controllers generally work better than controllers of higher degrees, while for higher degree systems, PWA controllers do not work well.
There are some limitations. 
First, the controller synthesis process is heuristic, providing no stability guarantees for the closed-loop system.
Controllable regions have to be computed a posteriori.
Second, besides the degree of the controller, there is not much more room for parameter tuning.
One possible way to tune parameters is to change the state space constraints that are not hard imposed.
For example, increasing or decreasing the limit on the maximum velocity of a rigid body would result in different controllers. 
Third, if the truncated moments in the objective of the SDP (\ref{sdp_dual}) are large, the SDP solver might run into numerical issues, due to the immaturity of the current SDP solvers and the Spotless software. 
A possible solution is to rescale the state space constraints together with the system dynamics.

\bibliographystyle{IEEEtran}
\bibliography{references}

\end{document}